\documentclass[aps,pre,twocolumn]{revtex4}
\usepackage{amsmath, amssymb, epsfig}

\renewcommand{\vr}{{\bf{r}}}
\newcommand{\vp}{{\bf{p}}}
\newcommand{\vq}{{\bf{q}}}
\newcommand{\vQ}{{\bf{Q}}}

\begin{document}

\title{Long-time saturation of the Loschmidt echo in quantum chaotic
  billiards}

\author{Martha Guti\'errez$^1$ and Arseni Goussev$^2$}

\affiliation{$^1$Institut f\"ur Theoretische Physik, Universit\"at
  Regensburg, 93040 Regensburg, Germany\\ $^2$School of Mathematics,
  University of Bristol, University Walk, Bristol BS8 1TW, United
  Kingdom}

\date{\today}

\begin{abstract}
  The Loschmidt echo (LE) (or fidelity) quantifies the sensitivity of the time
  evolution of a quantum system with respect to a perturbation of the
  Hamiltonian. In a typical chaotic system the LE has been previously argued
  to exhibit a long-time saturation at a value inversely proportional to the
  effective size of the Hilbert space of the system. However, until now no
  quantitative results have been known and, in particular, no explicit
  expression for the proportionality constant has been proposed. In this paper
  we perform a quantitative analysis of the phenomenon of the LE saturation
  and provide the analytical expression for its long-time saturation value for
  a semiclassical particle in a two-dimensional chaotic billiard. We further
  perform extensive (fully quantum mechanical) numerical calculations of the
  LE saturation value and find the numerical results to support the
  semiclassical theory.
\end{abstract}

\maketitle

\section{Introduction}

In his seminal 1984 paper \cite{peres}, Peres studied the stability of motion
of quantum systems with respect to small perturbations of the Hamiltonian. He
discovered that the quantum motion of a system, whose underlying classical
dynamics is chaotic, is more unstable than that of a system, whose dynamics is
regular in the classical limit. The quantity introduced by Peres, presently
known as the {\it Loschmidt echo} (LE) or {\it fidelity}, has been the subject
of thorough theoretical and experimental research in the fields of quantum
chaos and quantum information \cite{reviews, nielsen}.

The LE, defined as
\begin{equation}
  M(t) = |O(t)|^2
\label{eq1-01}
\end{equation}
with the amplitude
\begin{equation}
  O(t) = \langle \phi_0 | e^{i\tilde{H}t/\hbar} e^{-iHt/\hbar} | \phi_0
  \rangle \, ,
\label{eq1-02}
\end{equation}
quantifies the ``distance'' (in the Hilbert space) between the state
$e^{-iHt/\hbar} |\phi_0\rangle$, resulting form the initial state
$|\phi_0\rangle$ in the course of evolution through a time $t$ under the
Hamiltonian $H$, and the state $e^{-i\tilde{H}t/\hbar} |\phi_0\rangle$
obtained by evolving the same initial state through the same time $t$, but
under a slightly different, perturbed Hamiltonian $\tilde{H}$. The LE, by
construction, equals unity for $t=0$ and typically decays further in time. A
variety of different decay regimes -- the most prominent ones being the {\it
  Lyapunov} \cite{jalabert}, {\it Fermi-Golden-Rule} \cite{jacquod, jalabert},
and the {\it perturbative} \cite{jacquod, cerruti, prosen} regime -- have been
found in chaotic systems with various Hamiltonians and Hamiltonian
perturbations. In this paper, however, we address the property of the LE
generally shared by all (Hermitian) chaotic systems: the saturation of the
decay at long times.

Peres provided in his original work \cite{peres} a {\it qualitative}
(order-of-magnitude) estimate for the value $M_{\infty}$ of the LE saturation
in chaotic systems. He argued that for small enough perturbations
\begin{equation}
M_{\infty} \sim N^{-1} \, ,
\label{eq1-03}
\end{equation}
where $N$ is the number of eigenstates (of the unperturbed Hamiltonian $H$)
that are significantly represented in the initial state $| \phi_0 \rangle$. In
other words, $N$ is the size of the effective Hilbert space that is required
for a reasonable description of the time evolution of the initial state.

The phenomenon of the LE saturation has been previously addressed in the
literature from numerical \cite{cucchietti} and analytical \cite{petitjean}
perspective, and the validity of the Peres' argument, Eq.~(\ref{eq1-03}), has
been verified.  However, no explicit expression for the proportionality
constant in Eq.~(\ref{eq1-03}) has been proposed. Our work complements the
theory of the LE in chaotic systems by providing the (previously missing)
proportionality constant.

In this paper we present the semiclassical analysis of the LE at long times,
and derive an expression for the LE saturation value. Our result, while in
agreement with Eq.~(\ref{eq1-03}), constitutes a {\it quantitative} estimate
of $M_{\infty}$. The system treated in this paper is a two-dimensional,
quantum billiard that exhibits chaotic dynamics in the classical limit. The
key method underlying our analytical calculation, however, is not restricted
to billiards and can be generalized to a wider range of chaotic systems. We
further perform numerical simulations of the time evolution of an initially
localized, Gaussian wave packet in a chaotic billiard, and compute the
saturation of the LE due to a perturbation caused by a deformation of the
boundary. The results of the numerical simulation strongly support our
analytical predictions.  Finally, we conclude the paper with a discussion and
final remarks.

\section{Semiclassical approach}

We consider the time evolution of a quantum particle moving inside a
two-dimensional ballistic cavity -- a quantum billiard. In this paper we only
consider hard-wall billiards whose underlying classical dynamics is fully
hyperbolic \cite{ott}. The initial state of the particle is assumed to be the
coherent state
\begin{equation}
  \phi_0(\vr) = \frac{1}{\sqrt{\pi}\sigma} \exp\left[ \frac{i}{\hbar}
    \vp_0 \cdot (\vr-\vr_0) - \frac{(\vr-\vr_0)^2}{2\sigma^2} \right] \, ,
\label{eq2-01}
\end{equation}
with $\sigma$ quantifying the dispersion of the wave packet, and $\vr_0$ and
$\vp_0$ representing respectively the average position and momentum of the
particle. The dispersion $\sigma$ is assumed to be small compared to the
linear size of the billiard for the wave function to be normalizable to unity.
We further define the de Broglie wavelength of the particle as
\begin{equation}
  \lambda = \frac{2\pi\hbar}{p_0} \, ,
\label{eq2-01.5}
\end{equation}
where $p_0 = |\vp_0|$ is the magnitude of the particle's momentum.
(Hereinafter we denote the magnitude of a vector by its corresponding symbol
in italics.)

The time evolution of the initial state in the unperturbed system with the
Hamiltonian $H$ is given by $\phi_t(\vr) = \int d\vr' K_t(\vr,\vr')
\phi_0(\vr')$. In the (short-wavelength) {\it semiclassical} approximation the
propagator $K_t(\vr,\vr')$, for a two-dimensional system, can be written as
\cite{Gutzwiller}
\begin{equation}
  K_t(\vr,\vr') \approx \frac{1}{2\pi i \hbar}\sum_{\gamma'(\vr' \to \vr, t)}
  D_{\gamma'} \, e^{ i S_{\gamma'} / \hbar } \, .
\label{eq2-01.7}
\end{equation}
Here, $S_{\gamma'}$ denotes the action integral along the {\it classical} path
$\gamma'$ leading from the position $\vr'$ to $\vr$ in time $t$, and
$D_{\gamma'}\!=\!\left|\det(-\partial^2S_{\gamma'} / \partial \vr \partial
  \vr' )\right|^{1/2} \!{\rm e}^{-i\pi\mu_{\gamma'}/2}$ with the Maslov index
$\mu_{\gamma'}$. Then, in the limit (see Appendix~A of Ref.~\cite{goussev-1})
\begin{equation}
  \lambda \ll 2\pi \sigma \ll \sqrt{2\pi \lambda l_\mathrm{L}} \, ,
\label{eq2-01.75}
\end{equation}
with $l_\mathrm{L}$ being the Lyapunov length of the billiard, the action
integral $S_{\gamma'}$ can be linearized about the trajectory $\gamma(\vr_0
\rightarrow \vr,t)$ connecting the wave packet center $\vr_0$ and the point
$\vr$ in time $t$: $S_{\gamma'} \approx S_{\gamma} -
\vp_{\gamma}^{(\mathrm{i})} \cdot (\vr'-\vr_0)$, where
$\vp_{\gamma}^{(\mathrm{i})}$ is the initial momentum on the trajectory
$\gamma$. Using this action linearization and performing a Gaussian
integration over the initial point $\vr'$ one obtains the semiclassical
expression for the time-dependent wave function evolving under $H$ \cite{jalabert}:
\begin{eqnarray}
  \phi_t(\vr) &\approx& \frac{\sigma}{\sqrt{\pi} i \hbar}
  \sum_{\gamma(\vr_0 \rightarrow \vr,t)} D_{\gamma}
  \nonumber\\ && \times \exp\left[ \frac{i}{\hbar} S_{\gamma} -
    \frac{\sigma^2}{2\hbar^2} (\vp_{\gamma}^{(\mathrm{i})}-\vp_0)^2
    \right] \, .
\label{eq2-01.8}
\end{eqnarray}
The wave function $\tilde{\phi}_t(\vr)$ corresponding to the time evolution
under the Hamiltonian $\tilde{H}$ of the perturbed system is given by an
equation analogous to Eq.~(\ref{eq2-01.8}) with the trajectories $\gamma(\vr_0
\rightarrow \vr,t)$ replaced by $\tilde{\gamma}(\vr_0 \rightarrow \vr,t)$
satisfying the classical evolution corresponding to $\tilde{H}$.

The LE amplitude, $O(t) = \langle \tilde{\phi}_t | \phi_t \rangle$, is given
by
\begin{eqnarray}
  O(t) &\approx& \displaystyle \frac{\sigma^2}{\pi\hbar^2} \int d\vr
  \sum_{\gamma,\tilde\gamma(\vr_0 \rightarrow \vr,t)} D_{\gamma}D_{\tilde{\gamma}}^{*}\exp\left[
    \frac{i}{\hbar} (S_{\gamma}-S_{\tilde{\gamma}}) \right] \nonumber\\
 &&\times \exp\left[- \frac{\sigma^2}{2\hbar^2}
    \left( (\vp_{\gamma}^{(\mathrm{i})}\!-\!\vp_0)^2 +
      (\vp_{\tilde{\gamma}}^{(\mathrm{i})}\!-\!\vp_0)^2 \right) \right]  .
\label{eq2-02}
\end{eqnarray}
The expression for the LE, then, being the product $O^*(t) O(t)$, with the
asterisk denoting the complex conjugation, involves two integrals over the
final points, say $\vr$ and $\tilde{\vr}$, over four sums over trajectories,
two corresponding to the perturbed system and two to the unperturbed one. The
integrand, in general, is a rapidly oscillating function of $\vr$ and
$\tilde{\vr}$; therefore, only the trajectories with the overall phase
difference smaller than $\hbar$ give a finite contribution to the integral.
Considering trajectories such that $S_{\gamma}\approx S_{\tilde{\gamma}}$
leads to exponentially decaying regimes of the LE \cite{jalabert}. Therefore,
non-decaying contributions to the LE (responsible for the LE saturation) can
only result from trajectories that are close in action and belong to the same
Hamiltonian. This imposes a restriction on the possible configurations of the
trajectories of interest, namely $\vr \approx \tilde{\vr}$. This makes it
convenient to make the following transformation to the new integration
coordinates: $\vQ = (\vr + \tilde{\vr})/2$ and $\vq = \vr - \tilde{\vr}$.
Then, following the procedure above, we linearize the four trajectories
entering the expression for the LE about the same final point $\vQ$ to obtain
\begin{eqnarray}
  \lefteqn{ M(t) \approx \displaystyle \frac{\sigma^4}{\pi^2\hbar^4} \int d\vQ
    \int d\vq \sum_{\gamma,\gamma',\tilde{\gamma},\tilde{\gamma}'}
    D_{\gamma} D_{\gamma'}^* D_{\tilde{\gamma}}^*
    D_{\tilde{\gamma}'} } \nonumber\\ && \times
  \exp\left\{ \frac{i}{\hbar}\Delta S -\frac{\sigma^2}{2\hbar^2} \!\!
    \sum_{\gamma_i = \{ \gamma,\gamma',\tilde{\gamma},\tilde{\gamma}' \} }
    \!\!\!\!\!\!\! (\vp_{\gamma_i}^{(\mathrm{i})}-\vp_0)^2   \right\} , 
\label{eq2-03}
\end{eqnarray}
where $\Delta
S=(S_{\gamma}-S_{\gamma'}-S_{\tilde{\gamma}}+S_{\tilde{\gamma}'})+(\vp_{\gamma}^{(\mathrm{f})}
+ \vp_{\gamma'}^{(\mathrm{f})} - \vp_{\tilde{\gamma}}^{(\mathrm{f})} -
\vp_{\tilde{\gamma}'}^{(\mathrm{f})}) \cdot \vq/2$, all the four paths
($\gamma$, $\gamma'$, $\tilde{\gamma}$ and $\tilde{\gamma}'$) connect $\vr_0$
and $\vQ$ in time $t$, with two of them ($\gamma$ and $\gamma'$) corresponding
to the unperturbed Hamiltonian $H$ and the other two ($\tilde{\gamma}$ and
$\tilde{\gamma}'$) to the perturbed Hamiltonian $\tilde{H}$; here
$\vp^{(\mathrm{f})}$ denotes the final momentum (at the end point $\vQ$) on
the corresponding classical path. The integrand in Eq.~(\ref{eq2-03}) is still
a rapidly oscillating function of $\vQ$ -- $\Delta S$ is generally much
greater than $\hbar$ and is sensitive to $\vQ$ -- unless the paths $\gamma$,
$\gamma'$, $\tilde{\gamma}$ and $\tilde{\gamma}'$ are correlated. In the {\it
  diagonal approximation} \cite{berry} the main contribution to the
$\vQ$-integral comes from such terms in the sum that
$S_{\gamma}-S_{\gamma'}-S_{\tilde{\gamma}}+S_{\tilde{\gamma}'} = 0$. One group
of such terms, defined by the identification $\gamma = \tilde{\gamma}$ and
$\gamma' = \tilde{\gamma}'$, is responsible for the (generally exponential)
time-decay of the LE \cite{jalabert}, and leads to a vanishing contribution at
long times. The other group, defined by the identification $\gamma = \gamma'$
and $\tilde{\gamma} = \tilde{\gamma}'$, gives rise to a term surviving in the
limit $t \rightarrow \infty$ and, therefore, provides the leading order
contribution to the LE saturation value.  Thus, identifying the trajectories
of the unperturbed ($\gamma = \gamma'$) and perturbed ($\tilde{\gamma} =
\tilde{\gamma}'$) Hamiltonian we obtain
\begin{eqnarray}
  M_{\infty} &\approx& \frac{\sigma^4}{\pi^2\hbar^4} \int d\vQ \int d\vq
  \sum_{\gamma,\tilde{\gamma}} |D_{\gamma}|^2 |D_{\tilde{\gamma}}|^2
  \nonumber\\ && \times \exp\left\{ \frac{i}{\hbar}
    (\vp_{\gamma}^{(\mathrm{f})} - \vp_{\tilde{\gamma}}^{(\mathrm{f})})
    \cdot \vq \right. \nonumber\\ &&\left. -\frac{\sigma^2}{\hbar^2} \left[
      (\vp_{\gamma}^{(\mathrm{i})}-\vp_0)^2 +
      (\vp_{\tilde{\gamma}}^{(\mathrm{i})}-\vp_0)^2 \right] \right\} \, .
\label{eq2-04}
\end{eqnarray}

In order to evaluate the double sum in the right hand side of
Eq.~(\ref{eq2-04}) we utilize the sum rule \cite{argaman}
\begin{eqnarray}
  \lefteqn{ \sum_{\gamma(\vr \rightarrow \vr',t)} |D_{\gamma}|^2 f(\vr,
    \vp_{\gamma}^{(\mathrm{i})}; \vr', \vp_{\gamma}^{(\mathrm{f})}) }
  \nonumber\\ &&\phantom{+} = \int d\vp \int d\vp' \;
  \mathcal{P}_t(\vr, \vp; \vr', \vp') f(\vr, \vp; \vr', \vp') \, ,
\label{eq2-05}
\end{eqnarray}
where $\mathcal{P}_t(\vr, \vp; \vr', \vp') = \delta(\vr_{\gamma}(t)-\vr')
\delta(\vp_{\gamma}(t)-\vp')$ is the classical phase-space probability density
for a trajectory $\gamma = \left\{ (\vr_{\gamma}(\tau), \vp_{\gamma}(\tau)),
  \tau \in [0,t] \right\}$ starting from the phase-space point
$(\vr_{\gamma}(0), \vp_{\gamma}(0)) = (\vr,\vp)$ to end at the point
$(\vr_{\gamma}(t), \vp_{\gamma}(t)) = (\vr',\vp')$ while evolving under the
Hamiltonian $H$ through time $t$. Then, since dealing with chaotic
Hamiltonians and long times, we replace the probability distribution
$\mathcal{P}_t$ by its phase-space average
\begin{equation}
  \overline{\mathcal{P}}(\vr, \vp; \vr', \vp') = \frac{\delta(H(\vr',\vp') - 
    H(\vr,\vp))}{\Omega(H(\vr,\vp))},
\label{eq2-06}
\end{equation}
where $\Omega(E)$ is the phase-space volume of the energy shell
$H(\vr,\vp)=E$. For the case of two-dimensional billiards $\Omega(E) = 2\pi m
A$, with $m$ being the mass of the particle and $A$ the billiard area, so that
in view of Eqs.~(\ref{eq2-05}) and (\ref{eq2-06}) the long time ($t
\rightarrow \infty$) limit of Eq.~(\ref{eq2-04}) reads
\begin{eqnarray}
  \lefteqn{ M_{\infty} \approx  \frac{\sigma^4}{\pi^2\hbar^4\Omega^2}
    \int d\vQ \int d\vq }
  \nonumber\\ && \iiiint d\vp d\vp' d\tilde{\vp} d\tilde{\vp}' \,\delta(H(\vr_0,\vp)-H({\mathbf Q},\vp') )
  \nonumber\\ &&
  \times \delta(\tilde H(\vr_0,\tilde{\vp})-\tilde H({\mathbf Q},\tilde{\vp}')
  )  \exp \left\{ \frac{i}{\hbar} (\vp'-\tilde{\vp}') \cdot \vq \right.
  \nonumber\\ && \phantom{+++} \left. -\frac{\sigma^2}{\hbar^2} \left[
      (\vp-\vp_0)^2 + (\tilde{\vp}-\vp_0)^2 \right] \right\} \, .
\label{eq2-07}
\end{eqnarray}
We now assume that both Hamiltonians can be written as $\vp^2/2m+V(\vr)$ and
perform the integration in the right hand side of Eq.~(\ref{eq2-07}) as
follows. The $\vq$-integration, with the integration limits extended to
$\mathbb{R}^2$, results in $(2\pi\hbar)^2 \delta(\vp' - \tilde{\vp}')$.
Consequently integrating over $\vp'$ and $\tilde{\vp}'$ we obtain
\begin{eqnarray}
  \lefteqn{ M_{\infty}
    \approx \frac{8\pi m \sigma^{4}}{\hbar^{2}\Omega^2} \int d\mathbf{Q}
    \iint d\vp d\tilde{\vp} } \nonumber \\
  && \times
  \exp\left(-\frac{\sigma^{2}}{\hbar^{2}}\left[(\vp-\vp_{0})^{2}+(\tilde{\vp}-\vp_{0})^{2}\right] \right) \nonumber \\ && \times
  \delta(\Sigma(\vr_0)-\Sigma(\mathbf{Q})+(\vp^2/2m-\tilde{\vp}'^2/2m))
  \, ,
  \label{eq2-07.5}
\end{eqnarray}
where $\Sigma(\vr)=V(\vr)-\tilde V(\vr)$. Now we assume that the perturbation
is small compared to the kinetic part of the Hamiltonian. Alternatively, one
may consider perturbations of the Hamiltonian produced by deformations of the
billiard boundary \cite{goussev-1, goussev-2}; it is the perturbation of the
latter type that we use in our numerical experiments of the following section.
Thus, assuming $\Sigma = 0$, we have
\begin{eqnarray}
  \lefteqn{ M_{\infty} \approx \frac{4\sigma^2}{\pi\hbar^2 A} \iint d\vp
    d\tilde{\vp} \, \delta(p^2-\tilde{p}^2) } \nonumber\\ && \times \exp\left\{
    -\frac{\sigma^2}{\hbar^2} \left[ (\vp-\vp_0)^2 + (\tilde{\vp}-\vp_0)^2
    \right] \right\} \, .
\label{eq2-08}
\end{eqnarray}

Now, we use the integral representation of the $\delta$-function,
$\delta(p^2-\tilde{p}^2) = (2\pi)^{-1} \int d\xi \exp(i\xi p^2 - i\xi
\tilde{p}^2)$, and perform the Gaussian integration over $\vp$ and
$\tilde{\vp}$ (eventually doing the variable change $x=\xi\hbar^2/\sigma^2$)
to get
\begin{eqnarray}
  M_{\infty} &\approx& \frac{2\sigma^2}{A} \int_{-\infty}^{+\infty}
  \frac{dx}{1+x^2} \exp \left( -2a \frac{x^2}{1+x^2} \right) \nonumber\\
  && \phantom{+} = \frac{2\pi\sigma^2}{A} I_0(a) \exp(-a) \, ,
\label{eq2-09}
\end{eqnarray}
where $a = (p_0 \sigma / \hbar)^2 = (2\pi\sigma / \lambda)^2$, and $I_0$ is
the zeroth order modified Bessel function of the first kind. In the limit $a
\gg 1$ (or $\lambda \ll \sigma$), which is in agreement with
Eq.~(\ref{eq2-01.75}), the asymptotic form $I_0(a) \approx (2\pi a)^{-1/2}
\exp(a)$ yields
\begin{equation}
  M_{\infty} \approx \frac{1}{\sqrt{2\pi}} \frac{\lambda \sigma}{A} \, .
\label{eq2-10}
\end{equation}
Equation~(\ref{eq2-10}) constitutes the central analytical result of our
paper.

It is easy to see that the original argument by Peres, see Eq.~(\ref{eq1-03}),
is in perfect agreement with Eq.~(\ref{eq2-10}) derived in the semiclassical
approximation. Indeed, the number of Hamiltonian eigenstate required to
properly describe the time evolution of the initial wave packet, given by
Eq.~(\ref{eq2-01}), can be evaluated as $N = \Omega(E) \Delta E /
(2\pi\hbar)^2$. Here, as above, $\Omega(E) = 2\pi m A$ is the phase-space
volume of the energy shell at the average energy $E = p_0^2 / 2m$ of the
particle, and $\Delta E = p_0 \Delta p / m$ is the energy dispersion of the
initial state. Estimating the momentum dispersion as $\Delta p \approx
2\sqrt{2} \hbar / \sigma$ we obtain the following expression for the number of
the eigenstates: $N \approx 2\sqrt{2} A / \lambda \sigma$ (and therefore
$M_{\infty} \approx 2\pi^{-1/2} N^{-1}$). In fact, due to certain
arbitrariness in determination of $\Delta p$ the size of the effective Hilbert
space $N$ is not properly defined. This difficulty points to a drawback of the
original formulation of Eq.~(\ref{eq1-03}). On the contrary,
Eq.~(\ref{eq2-10}) gives the LE saturation value in terms of well defined
system parameters, $\lambda$, $\sigma$, and $A$, and, therefore, provides a
quantitative estimate for $M_{\infty}$.

In the following section we demonstrate that the semiclassical
predictions of Eq.~(\ref{eq2-10}) are in agreement with the time
saturation of the LE observed in numerical experiments.

\section{Numerical simulations}

In order to support our semiclassical calculations we have performed numerical
simulations of a quantum particle moving inside a desymmetrized diamond
billiard (DDB). The DDB is defined as a fundamental domain of the area
confined by four intersecting disks centered at the vertices of a square. The
billiard is fully chaotic \cite{kramli} and has been previously considered for
studying various aspects of quantum chaos \cite{goussev-1, goussev-2, muller}.
In our numerical experiments we used the piston-like boundary deformation
\cite{goussev-1} as the perturbation of the Hamiltonian. The numerical method
that we used for propagating the particle's wave function in time is the
Trotter-Suzuki algorithm \cite{raedt}; Reference~\cite{goussev-1} provides
further details on the billiard systems, Hamiltonian perturbation and wave
function time propagation.

\begin{figure}[h]
\centerline{\epsfig{figure=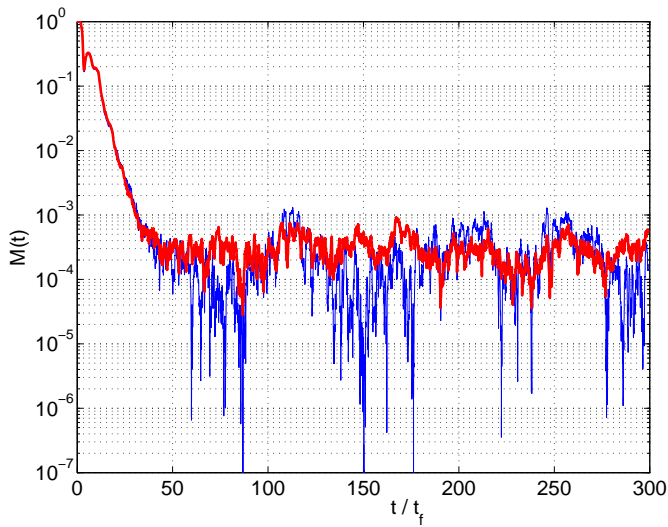,width=3.5in}}
\caption{(Color online) Time decay of the Loschmidt echo in the desymmetrized
  diamond billiard with a boundary deformation for initial wave packets of de
  Broglie wavelength $\lambda = 4\pi$ and dispersion $\sigma = 9$. Time is
  given in units of the free flight time $t_{\mathrm{f}}$ of the corresponding
  classical particle.  The thin (blue) line shows an individual LE decay curve
  resulted from a single numerical experiment. The thick (red) line represents
  the result of an averaging over 3 individual decay curves obtained for
  different positions $\vr_0$ of the initial wave packet.}
\label{fig-1}
\end{figure}

\begin{figure}[h]
\centerline{\epsfig{figure=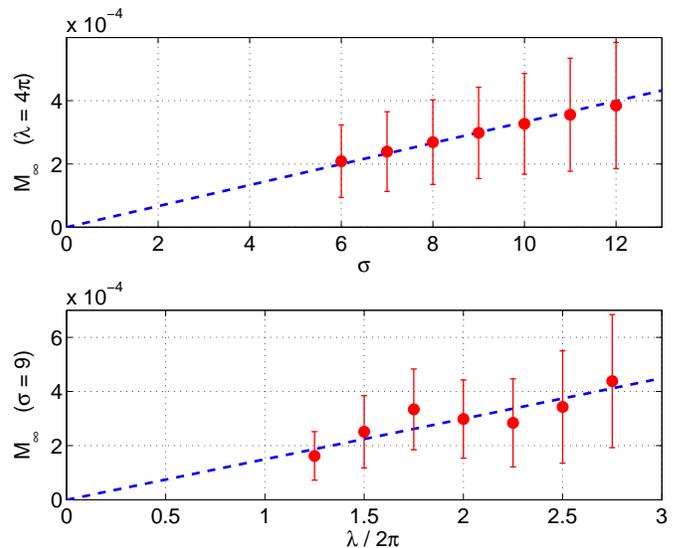,width=3.5in}}
\caption{(Color online) Top figure: The Loschmidt echo saturation value
  $M_{\infty}$ as a function of the dispersion $\sigma$ of the initial wave
  packet for a fixed de Broglie wavelength $\lambda = 4\pi$. Bottom figure:
  $M_{\infty}$ as a function of $\lambda$ for $\sigma = 9$. In both figures
  the billiard area $A \approx 1.51 \times 10^{5}$, and the blue dashed line
  represents the LE saturation value as predicted by Eq.~(\ref{eq2-10}).}
\label{fig-2}
\end{figure}

In our simulations the initial state of the quantum particle is given by
Eq.~(\ref{eq2-01}). The blue line in Fig.~\ref{fig-1} shows a typical LE decay
curve obtained in an individual numerical experiment with the initial wave
packet of the dispersion $\sigma = 9$ and de Broglie wavelength $\lambda =
4\pi$; the area of the billiard $A \approx 1.51 \times 10^{5}$. Time is given
in units of the free flight time $t_\mathrm{f}$ of the counterpart classical
billiard, i.e. $t/t_{\mathrm{f}}$ is the number of bounces of the
corresponding classical particle. The red line in Fig.~\ref{fig-1} is the
result of the averaging of the LE over 3 individual decay curves, each of
which was obtained by propagating a wave packet centered about a different
spatial point $\vr_0$ inside the billiard domain. (The wave packet centers
were chosen such that the three initial states had negligible overlap with one
another). Those were the average LE decay curves that we used to determine the
LE saturation value and standard deviation -- red dots and error bars in
Fig.~\ref{fig-2} -- for the initial quantum state with particular values of
the dispersion and de Broglie wavelength.

In Fig.~\ref{fig-2} we compare the semiclassical estimate for
$M_{\infty}$, given by Eq.~(\ref{eq2-10}), to the saturation values
obtained from the numerical simulations. The top (bottom) figure shows
the dependence of $M_{\infty}$ on the dispersion $\sigma$ (de Broglie
wavelength $\lambda$) for $\lambda = 4\pi$ ($\sigma = 9$) in the
billiard of the area $A \approx 1.51 \times 10^{5}$. The red dots
together with the error bars represent the numerically observed values
of $M_{\infty}$; the blue dashed lines are plotted in accordance with
Eq.~(\ref{eq2-10}). We stress here that no free (fitting) parameters
have been used in producing the theoretical lines: the slopes of the
lines are entirely fixed by Eq.~(\ref{eq2-10}). 

Finally, to give an idea of the scale of the numerical simulations of this
section we note that obtaining an individual LE decay curve, such as the blue
curve in Fig.~\ref{fig-1}, requires more than 8 days of computational time on
a high-end (2.8GHz, 2GB RAM) computer. Each data point in Fig.~\ref{fig-2} is a
result of the averaging over 3 such individual decay curves. Therefore, 39
individual decay curves were obtained to produce the numerical data presented
in Fig.~\ref{fig-2}, amounting to approximately 312 days of (single-processor)
computational time.

\section{Concluding remarks}

In this paper we have used the methods of the semiclassical theory to derive
an explicit expression for the value of the long-time saturation of the LE,
$M_{\infty}$, in two-dimensional chaotic billiards. Our quantitative result
agrees with the early qualitative argument \cite{peres} that the LE saturates
at a value inversely proportional to the effective size of the Hilbert space
of the system; our calculation provides the previously missing proportionality
factor.

In order to support our analytical predictions we have performed careful
numerical simulations of a quantum particle moving in a chaotic billiard. In
these simulations a deformation of the billiard boundary played the role a
Hamiltonian perturbation. The decay of the LE was observed until times long
enough to reliably determine $M_{\infty}$, and a proper ensemble averaging
(over the initial position of the quantum particle) was performed to improve
the accuracy. The numerically obtained values of the LE saturation were found
in a good agreement with the theory.

The central aspect of our semiclassical calculation is the pairing (in the
sense of the diagonal approximation) of trajectories that belong to the same
(perturbed or unperturbed) Hamiltonian. Those are these trajectory pairs that
render the time-independent contribution to the LE in addition to other,
exponentially decaying contributions resulting from different trajectory
pairs. Here we note that the pairing of trajectories considered in this work
has been previously studied in the context of the fidelity fluctuations
\cite{petitjean} and the survival probability decay in open chaotic systems
\cite{gutierrez}.

We also note that although the phenomenon of the long-time saturation of the
LE has been previously discussed in the literature \cite{petitjean,
  cucchietti} it has never been subject to a thorough analytical and/or
numerical study. In particular, the numerical simulations of the quantum
Lorentz gas \cite{cucchietti} correctly demonstrated the inverse
proportionality of $M_{\infty}$ to the billiard area, $A$, while misleadingly
suggesting its linear dependence on the square of the wave packet dispersion,
$\sigma^2$, along with independence of the de Broglie wavelength $\lambda$.
Reference~\cite{petitjean}, on the other hand, correctly outlined the
semiclassical derivation of the direct proportionality of $M_{\infty}$ to the
effective Plank constant, but did not present an explicit form of the
proportionality coefficient. Thus the present paper bridges the gap by
providing a quantitative analytical expression for the LE saturation value
and, consequently, verifying the expression by means of extensive numerical
simulations.

As the final remark we would like to point out that the present semiclassical
approach to the phenomenon of the LE saturation is only valid in the long time
limit and in the regime of weak Hamiltonian perturbations. In general,
however, the LE saturation value will depend on a (properly defined)
perturbation strength. It is not yet clear to us how this dependence can be
described by the semiclassical theory.

\section*{Acknowledgments}

The authors are thankful to Rodolfo Jalabert and Klaus Richter for helpful
discussions. MG acknowledges financial support from Deutsche
Forschungsgemeinschaft (Grant No. FG 760). AG acknowledges EPSRC for support under Grant
No. EP/E024629/1.

\end{document}